\documentclass[12pt,a4paper]{article}
\usepackage{amsfonts,latexsym}
\usepackage{graphicx,color}
\usepackage{dcolumn}
\usepackage{graphicx}
\usepackage{amsmath}
\usepackage{amsfonts}
\usepackage{amssymb}

\renewcommand{\thefootnote}{\fnsymbol{footnote}}

\begin{document}

\vspace{12mm}

\begin{center}
{{{\Large {\bf Slowly rotating black holes and their scalarization}}}}\\[10mm]

{Yun Soo Myung$^a$\footnote{e-mail address: ysmyung@inje.ac.kr} and De-Cheng Zou$^{b}$\footnote{e-mail address: dczou@yzu.edu.cn}}\\[8mm]

{${}^a$Institute of Basic Sciences and Department  of Computer Simulation, Inje University Gimhae 50834, Korea\\[0pt] }

{${}^b$Center for Gravitation and Cosmology and College of Physical Science and Technology, Yangzhou University, Yangzhou 225009, China\\[0pt]}
\end{center}
\vspace{2mm}

\begin{abstract}
We study scalarization  of slowly rotating  black holes
in the  Einstein-scalar-Gauss-Bonnet (GB)-Chern-Simons (CS) theory. In the slow rotation approximation of $a\ll1$ with rotation parameter $a$,
the GB term is given by a term  for Schwarzschild black hole,  whereas the CS term  takes a linear term of $a$.
The tachyonic instability for slowly rotating  black holes represents the onset of spontaneous scalarization.
We use  the (2+1)-dimensional  hyperboloidal foliation method to show  the tachyonic instability for slowly rotating  black holes by considering the time evolution of a spherically symmetric scalar mode. A threshold (existence) curve  is obtained  from the constant scalar modes under time evolution, which means the boundary between  stable and unstable black holes.
It is found  that
the slowly rotating black holes turn out to be  unstable against a spherically symmetric  scalar-mode
propagation for  positive coupling $\alpha$.
However, we could not find tachyonic  instability  and any $a$-bound for scalarization for negative coupling $\alpha$.

\end{abstract}
\vspace{5mm}

\vspace{1.5cm}

\hspace{11.5cm}
\newpage
\renewcommand{\thefootnote}{\arabic{footnote}}
\setcounter{footnote}{0}

%%%% Introduction %%%%

\section{Introduction}
Kerr black holes were used to derive spontaneous scalarization  in  the Einstein-scalar-Gauss-Bonnet-scalar (EsGB) theory together with positive coupling~\cite{Cunha:2019dwb,Collodel:2019kkx}. The tachyonic instability for Kerr black holes is considered as the onset of spin-induced  spontaneous scalarization.
Here, it is interesting to note  that  the sufficiently high rotations ($a\ge 0.5$) suppresses spin-induced scalarization  since the GB term might become  negative  outside the outer horizon.

Contrastively, the sufficiently high rotations ($a$-bound of $a/M \ge 0.5$) enhance  spin-induced scalarization for Kerr black holes in the same theory with  negative coupling~\cite{Dima:2020yac}.    This $a$-bound  was found  analytically by considering an $l\to\infty$-scalar mode~\cite{Hod:2020jjy} and numerically by solving  the (2+1)-dimensional evolution equation~\cite{Doneva:2020nbb}.
Further, any instability was  not  triggered when  $a<0.5$ with $M=1$ in the same theory~\cite{Zhang:2020pko}.
Consequently,  the spin-induced scalarized  black holes were constructed for sufficiently high rotations in the EsGB theory with negative coupling~\cite{Herdeiro:2020wei,Berti:2020kgk}. We wish to mention again that  spin-induced  scalarization was realized through  scalar couplings to the GB term.
However, we have found the threshold curve for tachyonic instability  without $a$-bound when performing   the instability  analysis for Kerr black holes~\cite{Myung:2020etf} and slowly rotating black holes~\cite{Myung:2021ztl} in the Einstein-scalar-Chern-Simons (EsCS) theory with negative coupling. This implies that the presence of $a$-bound represents a feature of the GB term together with negative coupling.

At this stage, it is very curious to introduce  slowly rotating black holes because they do not allow for sufficiently high rotations of $a\ge 0.5$.
These black holes could be  obtained  by confining  all quantities of interest  to first order in $a$ (that is, slow rotation approximation $a\ll1)$.
We would like to stress that
most black holes are born very slowly rotating~\cite{Fuller:2019sxi}. For example, black holes born from single stars rotate
very slowly  with $a=0.01$ and  fairly slow rotating black holes born from single stars  are regarded as  those   with $a\le0.1$.
So, we expect that scalarization of  slowly rotating black holes does not require $a$-bound, compared to the spin-induced scalarization of  Kerr black hole in the  EsGB theory with negative coupling.

In this work, we will investigate the onset of scalarization  for  slowly rotating  black holes
in the  Einstein-scalar-Gauss-Bonnet-Chern-Simons  (EsGBCS) theory with the scalar coupling parameter $\alpha$.  This theory includes  two single terms: the GB term being independent of $a$ and the CS term depending on $a\cos\theta$.  For our purpose, we need to introduce a setup of the numerical method.
We will use    the (2+1)-dimensional  hyperboloidal foliation method to derive  the tachyonic instability of slowly rotating  black holes for positive coupling $\alpha$  when choosing a spherically symmetric scalar-mode propagation. On the other hand, it may imply no tachyonic instability and thus, no $a$-bound for spontaneous scalarization when considering negative coupling $\alpha$.

\section{EsGBCS theory for slowly rotating black holes }
We start with the EsGBCS theory given by
\begin{eqnarray}
S=\frac{1}{16 \pi}\int d^4 x\sqrt{-g} \Big[
R-\frac{1}{2}(\partial \phi)^2+ \alpha\phi^2(R^2_{\rm GB}+{}^{*}RR)\Big]. \label{Action}
\end{eqnarray}
Here we use   geometric units of $G=c=1$.
We wish to describe briefly a significance of our action (\ref{Action}).
An action including both topological terms with different linear
couplings was firstly obtained from some superstring models~\cite{Antoniadis:1993jc} and the
heterotic strings~\cite{Cano:2019ore}.  Also, we have studied  spontaneous scalarization for Schwarzschild black hole in the action including
 both topological invariants with different quadratic couplings~\cite{Myung:2019wvb}.
Inspired by these, we introduced the action (\ref{Action}).

In Eq. (\ref{Action}),  the same  quadratic scalar coupling function  is chosen for two topological terms:
the GB term
\begin{equation} \label{gb1}
R^2_{\rm GB}=R^2-4R_{\mu\nu}R^{\mu\nu}+R_{\mu\nu\rho\sigma}R^{\mu\nu\rho\sigma}
\end{equation}
and the CS term
\begin{equation}
{}^{*}RR=\frac{1}{2}\epsilon^{\mu\nu\rho\sigma}R^{\eta}_{~\xi\rho\sigma}R^\xi_{~\eta\mu\nu}. \label{cs1}
\end{equation}
Varying  (\ref{Action}) with respect to  $g_{\mu\nu}$ and $\phi$ implies
 Einstein and scalar equations
\begin{eqnarray}
&&G_{\mu\nu}=\frac{1}{2}\partial_{\mu}\phi\partial_{\nu}\phi- \frac{1}{4}g_{\mu\nu}(\partial \phi)^2-4\alpha [\nabla^\rho \nabla^\sigma (\phi^2)P_{\mu\rho\nu\sigma}+ C_{\mu\nu}],\label{eqn-1}\\
&&\nabla^2\phi+2\alpha (R^2_{\rm GB}+{}^{*}RR)\phi=0,\label{eqn-2}
\end{eqnarray}
where $P_{\mu\rho\nu\sigma}$-tensor is a divergence-free part
of  Riemann tensor satisfying $\nabla_\mu P^{\mu}_{~\rho\nu\sigma}=0$.
Also,  $C_{\mu\nu}$ is the  Cotton tensor  given by
\begin{eqnarray}\label{cotton}
C_{\mu\nu}=\nabla_{\rho}(\phi^2)~\epsilon^{\rho\sigma
\gamma}_{~~~~(\mu}\nabla_{\gamma}R_{\nu)\sigma}+\frac{1}{2}\nabla_{\rho}\nabla_{\sigma}
(\phi^2)~\epsilon_{(\nu}^{~~\rho \gamma
\delta}R^{\sigma}_{~~\mu)\gamma \delta}.
\end{eqnarray}

Choosing $\phi=0$, Eq. (\ref{eqn-1}) reduces to $R_{\mu\nu}=0$ which allows  the  Kerr spacetime as a solution written in Boyer-Lindquist coordinates
\begin{eqnarray}
ds_{\rm Kerr}^2&\equiv&\tilde{g}_{\mu\nu}dx^\mu dx^\nu \nonumber \\
&=& -\frac{\Delta}{\rho^2}(dt-a \sin^2 \theta d\varphi)^2+\frac{\rho^2}{\Delta}dr^2
+\rho^2 d\theta^2 +\frac{\sin^2 \theta}{\rho^2}[a dt -(r^2+a^2)d\varphi]^2.   \label{kerr-st}
\end{eqnarray}
Here, we have mass $M$, angular momentum $J$,  rotation parameter $a=J/M>0$, $\Delta=r^2-2Mr +a^2$, and $\rho^2=r^2+a^2 \cos^2\theta$.
It is meaningful to note that Eq. (\ref{kerr-st}) describes a stationary, axisymmetric, and non-static  spacetime.
Considering $\Delta=0$ leads to the outer/inner horizons  as
$\tilde{r}_\pm=M\pm \sqrt{M^2-a^2}$.

In this case, two topological terms for Eq. (\ref{kerr-st}) are given by
\begin{eqnarray}
&&\tilde{R}^2_{\rm GB}=\frac{48M^2(r^6-15r^4 a^2\cos^2\theta+15 r^2a^4\cos^4\theta-a^6\cos^6\theta)}{\rho^{12}} \nonumber \\
&&\quad\quad \simeq \frac{48M^2}{r^6}\Big[1-\frac{21a^2\cos^2\theta}{r^2}+\cdots\Big],  \label{GB2} \\
&&{}^{*}\tilde{R}\tilde{R}=\frac{96rM^2a\cos \theta(3r^4-10r^2a^2\cos^2\theta+3a^4\cos^4\theta)}{\rho^{12}} \nonumber \\
  &&\quad\quad \simeq  \frac{96M^2 a \cos\theta}{r^7} \Big[3-\frac{28a^2\cos^2\theta}{r^2}+\cdots\Big].             \label{CS2}
\end{eqnarray}
Here $\tilde{R}^2_{\rm GB}({}^{*}\tilde{R}\tilde{R})$ is even (odd) with respect to parity transformation: $\tilde{R}^2_{\rm GB}(\pi-\theta)=\tilde{R}^2_{\rm GB}(\theta)$ and
${}^{*}\tilde{R}\tilde{R}(\pi-\theta)=-{}^{*}\tilde{R}\tilde{R}$. This  property plays an important role in conjecturing the threshold curve for negative coupling $\alpha$
together  with the transformation of $\alpha\to-\alpha$. Also,  the second term in Eq. (\ref{GB2}) implies that for negative $\alpha$, a spherical scalar mode is stable at low rotations ($a<0.5$) whereas  it is unstable at sufficiently high rotations ($a\ge 0.5$),  implying the $a$-bound of $a\ge 0.5$ with $M=1$ for spin-induced  scalarization in the EsGB theory~\cite{Dima:2020yac}.

Taking the slow rotation approximation ($a\ll 1$), we  introduce the slowly rotating black hole  keeping up to ${\cal O} (a)$-order~\cite{Lense:1918zz,Lammerzahl:2018zvb}
\begin{eqnarray}
ds_{\rm SRBH}^2&=&\bar{g}_{\mu\nu}dx^{\mu}dx^{\nu} \nonumber \\
&=& -\Big(1-\frac{2M}{r}\Big)dt^2+\frac{dr^2}{1-\frac{2M}{r}}
+r^2 (d\theta^2 +\sin^2 \theta d\varphi^2)+\frac{4aM \sin^2\theta}{r} dt d\varphi,   \label{srbh}
\end{eqnarray}
where the last term represents the axisymmetric and non-static spacetime.
Hereafter, we  ignore all terms involving higher order than $a$  in all other quantities of interest such that  $\bar{R}\simeq0,~\bar{R}_{\mu\nu}\simeq0,~ \bar{R}_{\mu\nu\rho\sigma}\not=0,~\cdots$.
In this case, the (outer) horizon is   given by the Schwarzschild radius as
\begin{equation}
r_+=2M,
\end{equation}
whereas the inner horizon ($r=r_-$) disappears.
Up to ${\cal O} (a)$-order,  two topological terms are  given by two single terms as
\begin{eqnarray}
&&\bar{R}^2_{\rm GB}\simeq  \frac{48M^2}{r^6}  \label{gb2}
\end{eqnarray}
and
\begin{eqnarray}
&&{}^{*}\bar{R}\bar{R}\simeq  \frac{288 M^2 a \cos\theta}{r^7}.       \label{cs2}
\end{eqnarray}
We wish to mention  that
Eq. (\ref{gb2}) is just the term for the Schwarzschild  black hole, while Eq. (\ref{cs2}) is a linear term which  approaches zero as $a\to 0$.
It clear from Eq. (\ref{gb2}) that  there is no  unstable spherical scalar mode
of $l=m=0$ for negative coupling $\alpha$ because the second term of $21a^2\cos^2\theta/r^2$ in Eq. (\ref{GB2}) disappears.  This might  imply no $a$-bound of $a\ge 0.5$ appeared in the Kerr black hole   in the EsGB theory with negative coupling $\alpha$.
On the other hand, considering Eq. (\ref{cs2}), the instability of the spherical mode is determined mainly by $a\cos \theta$.
This means that the rotation $a$   plays  a role in determining the instability of slowly rotating black holes, implying $\alpha>\alpha_{\rm th}(a)$-bound for positive coupling $\alpha$.  These indicate different  features of two topological terms when making the slow rotation approximation.
In the slow rotation approximation, the GB term provides the property
of the static solution, while the CS term gives
the property of non-static solution. This explains
that the scalarized Schwarzschild (slowly rotating) black holes could be found from the GB (CS)
coupling, but these are never found from the CS (GB) coupling.

Before we proceed, let us briefly  mention  the tachyonic instability on the non-rotating black holes.
If $a=0$, Eq. (\ref{srbh}) reduces to the line element for Schwarzschild black hole (SBH)
\begin{eqnarray}
ds_{\rm SBH}^2&=&\bar{g}_{\mu\nu}dx^{\mu}dx^{\nu} \nonumber \\
&=& -\Big(1-\frac{2M}{r}\Big)dt^2+\frac{dr^2}{1-\frac{2M}{r}}
+r^2 (d\theta^2 +\sin^2 \theta d\varphi^2).   \label{Sbh}
\end{eqnarray}
In this background, one has a  linearized scalar equation
\begin{equation}
\Big(\bar{\nabla}^2_{\rm SBH}+\frac{96\alpha M^2}{r^6}\Big)\delta\phi=0.\label{lins-eq2}
\end{equation}
Considering
\begin{equation} \label{scalar-sp}
\delta \phi(t,r,\theta,\varphi)=\frac{u(r)}{r}e^{-i\omega t}Y_{lm}(\theta,\varphi),
\end{equation}
and introducing a tortoise coordinate $r_*=r+2M\ln(r/(2M)-1)$ defined by $dr_*=dr/(1-2M/r) $, the radial equation of (\ref{lins-eq2}) leads to the Schr\"{o}dinger-type equation
\begin{equation}
\frac{d^2u}{dr_*^2}+\Big[\omega^2-V_{\rm SBH}(r)\Big]u(r)=0,
\end{equation}
where the potential $V_{\rm SBH}(r)$ is given by
\begin{equation} \label{pot-c}
V_{\rm SBH}(r)=\Big(1-\frac{2M}{r}\Big)\Big[\frac{2M}{r^3}+\frac{l(l+1)}{r^2}-\frac{96\alpha M^2}{r^6}\Big].
\end{equation}
In the case of $s(l=0)$-mode, from $\int^\infty_{2M} dr V_{\rm SBH}(r)/(1-2M/r)<0$,
we  obtain  a sufficient  condition of an unstable bound on  the  coupling  parameter $\alpha$~\cite{Doneva:2017bvd}
\begin{equation}
\alpha_{\rm sc}^{l=0}>0.4167M^2. \label{mass-b}
\end{equation}
However, (\ref{mass-b}) is not a  necessary and sufficient condition for instability.
In order to determine the tachyonic  instability precisely, one has to solve the second-order differential equation numerically
\begin{equation}\label{pertur-eq}
\frac{d^2u}{dr_*^2}-\Big[\Omega^2+V_{\rm SBH}(r)\Big]u(r)=0,
\end{equation}
which may  allow an exponentially growing mode of  $e^{\Omega t}(\omega=i\Omega) $ as  an unstable mode.
Here we choose two boundary conditions: a normalizable
solution of $u(\infty)\sim e^{-\Omega r_*}$  at infinity  and
a solution of $u(2M)\sim \left(r-2M\right)^{2M\Omega}$  near the horizon.
Actually, the bound for tachyonic  instability  is less than (\ref{mass-b}) and  given by $\alpha\ge 0.3628$ with $M=1$.
This shows that small $\alpha<0.3628$ is not enough to trigger the tachyonic instability.
The unstable region will be represented by a red-line on the $\alpha$-axis in the GBCS-threshold curve. The spontaneous scalarization of SBH was studied in the EsGB theory
~\cite{Doneva:2017bvd,Silva:2017uqg,Antoniou:2017acq} and the EsGBCS theory~\cite{Myung:2019wvb}.

Finally, we would like to mention that  the $s(l=0)$-mode analysis is not enough to deduce stability/instability because the analysis for higher mode with $l\ge1$
is necessary to be performed. In this case, from  (\ref{pot-c}), there exists a positive contribution of $l(l+1)/r^2$ to the potential with $l=0$.
It is worth noting that $V_{\rm SBH}$ with  $l\ge1$ indicates  potential barrier [see (Left) Fig. 1] for fixed  $\alpha=0.6$, whereas
it  shows potential well in the near horizon [see (Right) Fig. 1]  with $\alpha \ge 0.6$ for fixed $l$.  This implies that the $l=1$-mode may be stable even for the unstable region ($\alpha\ge0.3628$)  of the $l=0$ mode. We note that the shapes of potential  $V_{\rm SBH}$ with $l=0$ for $\alpha\le0.5$ are similar to (Right) Fig. 1.
 Analytically, the sufficient condition for instability could be derived  from $\int^\infty_{2M} dr V_{\rm SBH}(r)/(1-2M/r)<0$ as
\begin{equation}
\alpha_{\rm sc}^{l}> 0.4168 + 0.8333 l(l+1), \label{mass-g}
\end{equation}
which  means that increasing $l$ makes increasing $\alpha$-bound.
Explicitly, it implies for $l=1$
\begin{equation}
\alpha^{l=1}_{\rm sc}> 2.0833,\label{mass-l1}
\end{equation}
which is large than  the lowest bound of  $l=0$ given by  (\ref{mass-b}).

\begin{figure*}[t!]
   \centering
  \includegraphics{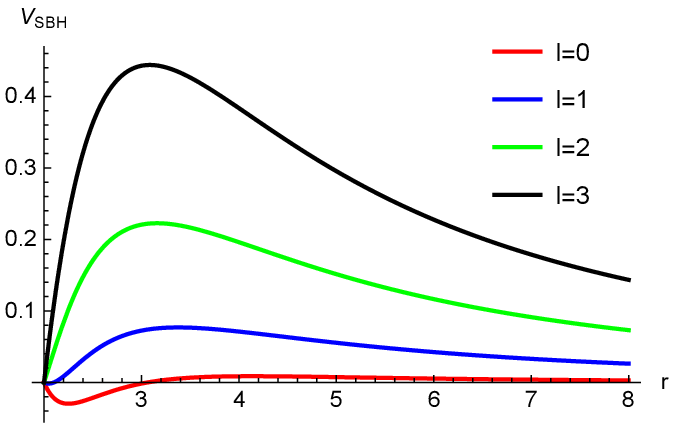}
  \hfill%
  \includegraphics{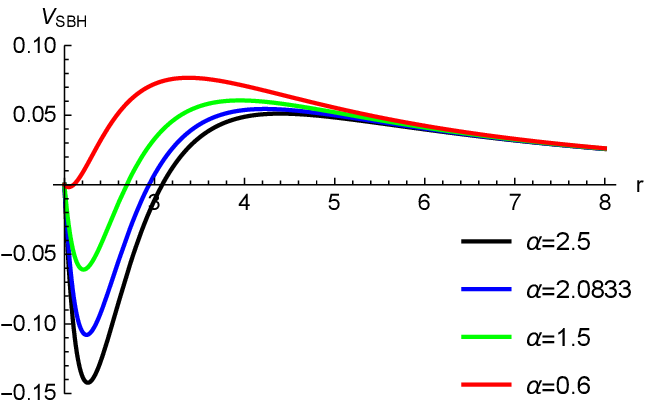}
\caption{Potentials  $V_{\rm SBH}$ around SBH: (Left) $V_{\rm SBH}$ with different $l$ for fixed $\alpha=0.6$.    (Right)  $V_{\rm SBH}$ with different $\alpha$ for fixed $l=1$, showing potential wells  in the near horizon. Here we choose $M=1$.   }
\end{figure*}

\section{Unstable slowly rotating black holes}
To explore the onset of scalarization for slowly rotating black holes,
the perturbations ($h_{\mu\nu},\delta \phi$) propagating
around the slowly rotating black hole background are introduced  as
\begin{eqnarray} \label{m-p}
g_{\mu\nu}=\bar{g}_{\mu\nu}+h_{\mu\nu},~~\phi=\bar{\phi}+\delta\phi \quad {\rm with}~ \bar{\phi}=0.
\end{eqnarray}
The linearized equation to (\ref{eqn-1}) takes a simple form like the general relativity as
\begin{eqnarray}\label{pertg}
\delta R_{\mu\nu}(h)\simeq 0,
\end{eqnarray}
where
\begin{eqnarray}\label{cottonp0}
\delta
R_{\mu\nu}(h)&=&\frac{1}{2}\left(\bar{\nabla}^{\gamma}\bar{\nabla}_{\mu}
h_{\nu\gamma}+\bar{\nabla}^{\gamma}\bar{\nabla}_{\nu}
h_{\mu\gamma}-\bar{\nabla}^2h_{\mu\nu}-\bar{\nabla}_{\mu} \bar{\nabla}_{\nu} h\right).\label{cottonp1}
\end{eqnarray}
The linearized scalar equation is important to test the stability of slowly rotating black holes and it  is given by
\begin{eqnarray}
 \Big(\bar{\nabla}^2-\mu^2_{\rm eff}\Big)\delta\phi=0\label{phi-eq2}
\end{eqnarray}
with an effective mass composed of two terms
\begin{equation} \label{eff-m}
\mu^2_{\rm eff}\equiv\mu^2_{\rm GB}+\mu^2_{\rm CS}=-2\alpha \bar{R}^2_{\rm GB}-2\alpha~{}^{*}\bar{R}\bar{R}(a).
\end{equation}
It is helpful to note that  a tensor-stability analysis for the slowly rotating black hole with Eq. (\ref{pertg}) is the same as in the general relativity.
One might not find  unstable tensor modes propagating around the slowly rotating black hole background~\cite{Hafner:2019kov}.
Hence, the stability of slowly rotating  black hole will be determined totally by the linearized scalar equation (\ref{phi-eq2}) in the EsGBCS theory.
We remind the reader that $\mu^2_{\rm GB}(\mu^2_{\rm CS})$
is variant (invariant)  under a combined transformation of $\alpha\to -\alpha$ and $\theta\to\pi-\theta$~\cite{Zhang:2021btn}.
From now on, we consider the case  of $\alpha>0$.
\begin{figure*}[t!]
   \centering
  \includegraphics{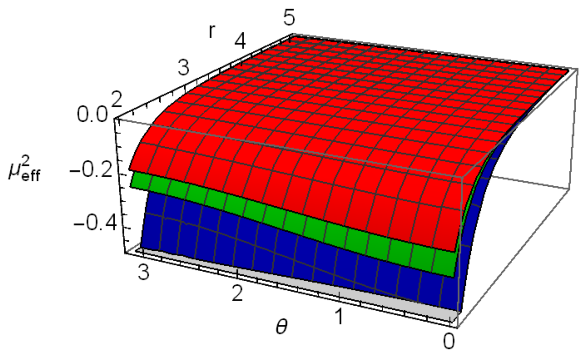}
  \hfill%
  \includegraphics{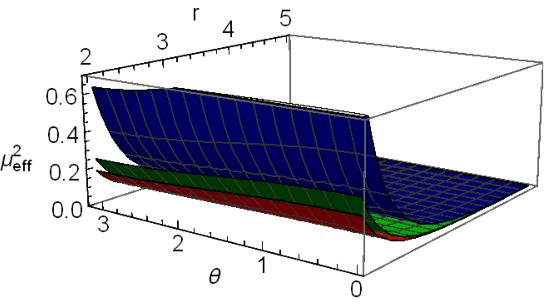}
\caption{3D graphs of $ \mu^2_{\rm eff}$ for  $a$=0.05 with positive $\alpha$ (Left) and $a$= 0.05 with negative $\alpha$ (Right) including  $r\in[r_+=2,5]$ and $\theta\in[0,\pi]$. The red, green, and blue surfaces represent $\alpha=\pm0.3,~ \pm0.4$ and $\pm1.0$, respectively. For positive $\alpha$ (Left), the potential wells
are deeper and deeper as $\alpha$ increases while the potential barriers are higher and higher as $-\alpha$ increases for negative $\alpha$ (Right).   }
\end{figure*}

We find from Eqs.  (\ref{gb2}) and (\ref{cs2}) that $\bar{R}^2_{\rm GB}$  is always positive and decreasing with $r$ (being independent of $a$), while  ${}^{*}\bar{R}\bar{R}(a)$  is an odd function with respect to $\cos\theta$. We observe from (Left) Fig. 2 that  $ \mu^2_{\rm eff}$ indicates  negative  regions outside the horizon.
The potential wells are deeper and deeper as $\alpha$ increases. It seems that  these may provide the tachyonic instability.
At this stage , we note that small negative $\mu^2_{\rm eff}|_{\alpha=0.3}$ is not sufficient to trigger the tachyonic instability whereas large negative $\mu^2_{\rm eff}|_{\alpha=1}$ is  sufficient to trigger the tachyonic instability.
On the other hand, from (Right) Fig. 2, the negative $\alpha$ case suggests  positive  regions outside the horizon, implying no instability.
However, the above  is  a rough estimation to see the instability even for positive $\alpha$ because $\mu^2_{\rm eff}$ depends $\alpha$ as well as $a$.
We note  that  the threshold  curve $\alpha_{\rm th}(a)$ for  slowly rotating  black holes depends on  $a$. It  will be determined precisely by carrying out numerical computations. As an example, after computing the  threshold of instability $\alpha_{\rm th}(a=0.05)=0.3627$, we know that it is stable  for $\alpha(=0.3)<\alpha_{\rm th}$ while it is unstable for  $\alpha(=0.4,1)>\alpha_{\rm th}$.

Let us briefly mention the (2+1)-dimensional hyperboloidal foliation method to solve Eq. (\ref{phi-eq2}) numerically~\cite{Gao:2018acg}.
Firstly, we  introduce the ingoing Kerr-Schild coordinates $\{\tilde{t},r,\theta,\tilde{\varphi}\}$ by considering the coordinate transformations
\begin{equation}
d\tilde{t}=dt+\frac{2 Mr}{\Delta} dr,\quad d\tilde{\varphi}=d\varphi +\frac{a}{\Delta} dr.
\end{equation}
Considering separation of variables
 \begin{equation}
 \delta \phi(\tilde{t},r,\theta,\tilde{\varphi}) =\frac{1}{r} \sum_{m} u_m(\tilde{t},r,\theta) e^{i m \tilde{\varphi}},
\end{equation}
Eq.(\ref{phi-eq2}) leads to a (2+1)-dimensional Teukolsky equation  as
\begin{equation}
A^{\tilde{t}\tilde{t}}\partial_{\tilde{t}}^2u_m+A^{\tilde{t}r}\partial_{\tilde{t}}\partial_ru_m+A^{rr}\partial^2_r u_m+
A^{\theta\theta}\partial_\theta^2u_m+B^{\tilde{t}}\partial_{\tilde{t}}u_m
+B^r\partial_r u_m+B^\theta\partial_\theta u_m+C u_m=0\label{phi-eq4}
\end{equation}
with coefficients
\begin{eqnarray}
&&A^{\tilde{t}\tilde{t}}=\rho^2+2Mr,~~A^{\tilde{t}r}=-4Mr,~~A^{rr}=-\Delta,~~A^{\theta\theta}=-1,\nonumber\\
&&B^{\tilde{t}}=2M,~~B^r=\frac{2}{r}(a^2-Mr)-2ima,~~B^\theta=-\cot\theta,  \label{coeffs}\\
&&C=\frac{m^2}{\sin^2\theta}-\frac{2(a^2-Mr)}{r^2}+\frac{2ima}{r}+\mu^2_{\rm eff}\rho^2. \nonumber
\end{eqnarray}
As the second step, we wish to solve Eq. (\ref{phi-eq4}) by adopting the hyperboloidal foliation method~\cite{Racz:2011qu}
with   compactified horizon-penetrating hyperboloidal (HH) coordinates $\{\tau, \rho,\theta, \tilde{\varphi}\}$.
In this case, Eq. (\ref{phi-eq4}) could be rewritten as
\begin{equation}
\partial^2_{\tau} u_m=\tilde{A}^{\tau\rho}\partial_{\tau} \partial_{\rho} u_m+\tilde{A}^{\rho\rho}\partial^2_\rho u_m+
\tilde{A}^{\theta\theta}\partial_\theta^2u_m+\tilde{B}^{\tau}\partial_{\tau}u_m
+\tilde{B}^\rho\partial_\rho u_m+\tilde{B}^\theta\partial_\theta u_m+\tilde{C} u_m=0, \label{phi-eq5}
\end{equation}
where coefficients appeared in~\cite{Zhang:2020pko}.
Finally, introducing a momentum $\Pi_m$,  one finds two coupled first-order equations as
\begin{eqnarray}
\partial_{\tau} u_m&=&\Pi_m, \label{phi-eq6} \\
\partial_{\tau} \Pi_m&=&\tilde{B}^{\tau} \Pi_m+
\tilde{A}^{\tau\rho}\partial_{\rho} \Pi_m +\tilde{A}^{\rho\rho}\partial_{\rho}^2 u_m+ \tilde{A}^{\theta\theta}\partial_{\theta}^2u_m+\tilde{B}^{\rho}\partial_{\rho} u_m+
\tilde{B}^{\theta}\partial_{\theta} u_m+\tilde{C} u_m. \label{phi-eq7}
\end{eqnarray}
\begin{figure*}[t!]
   \centering
  \includegraphics{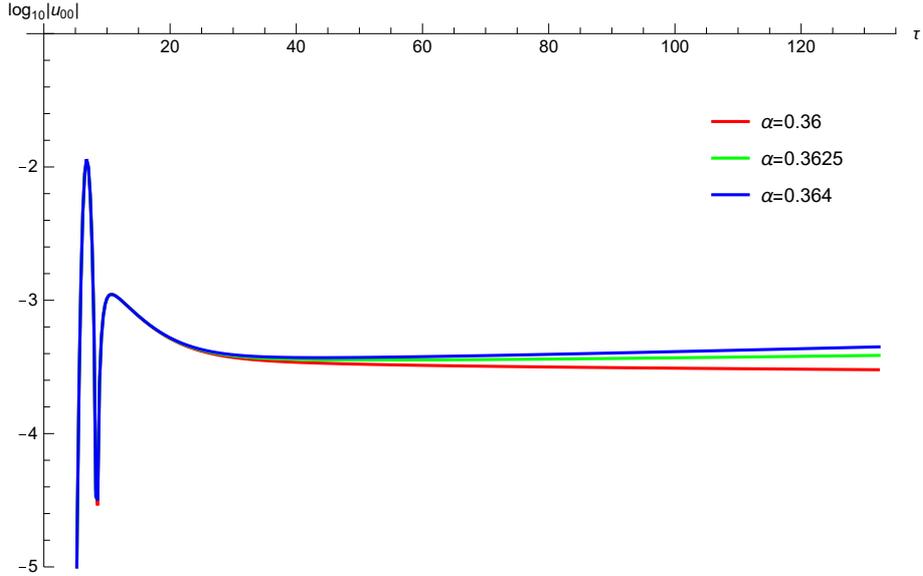}
\caption{Time evolution of  a scalar mode $\log_{10}|u_{00}(\tau,\rho=6M,\alpha)|$ for $a=0.15$ with three different  $\alpha$.
$\alpha=0.36$ represents stability ($\searrow$), $\alpha=0.3625$ denotes threshold ($\longrightarrow$), and  $\alpha=0.364$ represents instability ($\nearrow$). The middle curve represents  the threshold point at $a=0.15$ in Fig. 4. }
\end{figure*}

The  $\rho$ and $\theta$-differential equations  are solved by using the finite difference method, whereas the time ($\tau$) evolution is obtained by applying
the fourth-order Runge-Kutta integrator.
Using the HH coordinates leads to the fact  that the ingoing (outgoing) boundary conditions at the  horizon (infinity)
are satisfied automatically. Furthermore, the boundary conditions at the poles are given as $u_m|_{\theta=0,\pi}=0$ for odd $m$ and
$\partial_{\theta}u_m|_{\theta=0,\pi}=0$ for even $m$.

One introduces   a Gaussian function [$u_{lm}(\tau=0,\rho,\theta)\sim Y_{lm}(\theta)e^{-\frac{(\rho-\rho_c)^2}{2\sigma^2}}$] localized at $\rho=\rho_c$ outside the  horizon as an initial data for $u_{lm}$.
In addition, we may have  an initial boundary condition of  $\Pi_{lm}(\tau=0,\rho,\theta)=0$ if  Eq. (\ref{phi-eq6}) holds at $\tau=0$.
Here, we take $\rho_c=6M$ with $M=1$. Observers are assumed to be located at $\rho=6M$ and $\theta=\pi/4$.

The mode coupling may occur because the slowly rotating spacetime is not spherically symmetric.
This implies that different $l$-modes with the same $m$ are not independent and thus,  coupled to each other. 
However, it is suggested  that  the $l=m$ will become dominant.
As an example, the late-time dominant mode is  the mode with $l=m=0$  among  the processes starting with $l=0,1,2$ and $m=0$~\cite{Gao:2018acg}.   
This means that the $l=m=0$ case  is always a dominant mode whatever the initial mode is chosen. 
Concerning the range for $a$, we confine ourselves to  $0\le a\le 0.3$ including fairly slow rotating black holes ($0\le a\le 0.1$)  because we are considering the slowly rotating black holes. Also, we expect that the GBCS-threshold curve $\alpha=\alpha_{\rm th}(a)$ being the boundary between stable and unstable black holes may exist around $\alpha=0.3628$ because $\bar{R}^2_{\rm GB}$ in (\ref{gb2}) is dominant in the slow rotating approximation.

Here, we take   a spherically symmetric mode of $l=m=0$ as an initial mode.  
As is shown Fig. 3, the time  evolution for $\log_{10}|u_{00}(\tau,\rho=6M,\alpha)|$ provides stability ($\searrow$), threshold ($\longrightarrow$), and instability ($\nearrow$) with increasing time ($\tau$).

From Fig. 4, we find  a threshold  curve   $\alpha=\alpha_{\rm th}(a)$  which is the boundary between  stable and unstable black holes
based on the constant scalar modes $[\log_{10}|u_{00}(\tau,\rho,\alpha)|\sim \longrightarrow]$.
We observe that the GBCS-threshold curve decreases very slowly as $a$ increases and it  hits the $\alpha$-axis at $\alpha=0.3628$ when $a=0$.
The region for fairly slow rotating black holes denotes $0<a\le 0.1$ and the upper limit is represented by a dashed line at $a=0.1$.
The unshaded lower region [$\alpha<\alpha_{\rm th}(a)$: no growing mode] represents stable slowly rotating black holes, while the shaded upper region [$\alpha\ge \alpha_{\rm th}(a)$: growing mode]  denotes the unstable slowly rotating black holes.
Also, Fig. 4 includes   stable ($\alpha<0.3628$) and unstable (red-line: $\alpha\ge0.3628$) Schwarzschild black holes on the $\alpha$-axis which is derived  from the GB term solely. If one uses either $l=m=1$ or $l=m=2$ as initial modes,
the stable region (unshaded lower region) will be increased, compared to the $l=m=0$ case. 
\begin{figure*}[t!]
   \centering
  \includegraphics{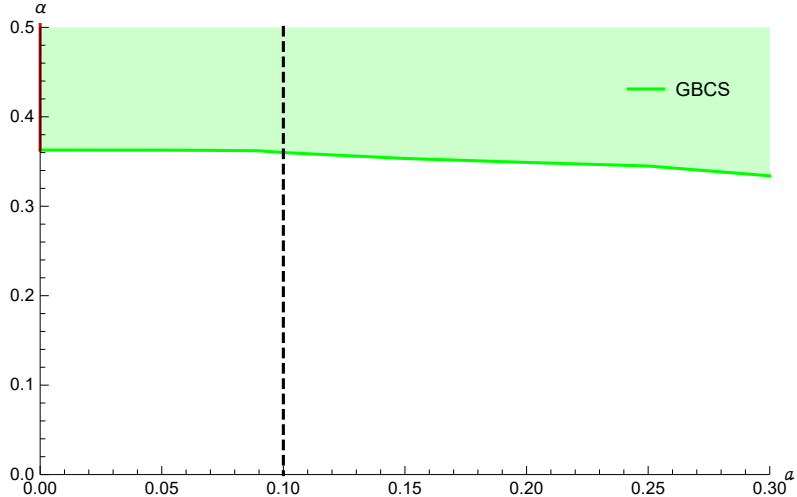}
\caption{GBCS-threshold (existence) curve $\alpha=\alpha_{\rm th}(a)$ being the boundary between stable  and unstable slowly  rotating black holes is obtained from observing time evolution of $l=m=0$-scalar mode  for positive $\alpha$.  The dashed line denotes the upper limit for fairly slow rotating black holes ($0<a\le 0.1$).
 Also, it includes  the stable and unstable (red-line: $\alpha\ge0.3628$) Schwarzschild black holes on the $\alpha$-axis found from the GB term only.
}
\end{figure*}

At this stage, we consider the case with negative coupling $\alpha$.
In this case, there is  no tachyonic instability for scalarization because $\mu^2_{\rm eff}$ has   positive region outside the horizon for negative $\alpha$ (see (Right) Fig. 2).  Also, this implies the absence of  $a$-bound for spontaneous scalarization.

Finally, we would like to mention the superradiant instability, which is known to exist at  high rotations for constant real  masses~\cite{Damour:1976kh,Zouros:1979iw,Detweiler:1980uk,Dolan:2007mj}. Further, superradiance appears if bosonic waves with angular momentum are amplified when
scattered by a spinning black hole, which spins down the black hole. Superradiant scattering could develop into an instability since the the  bosonic field is confined near the black hole by its constant mass. It is worth noting that  superradiant instability  may occur with nonconstant effective mass~\cite{Dima:2020rzg}.
In our work, however, the tachyonic instability is more plausible to occur than superradiant instability because  rapid falloffs of (\ref{gb2}) and (\ref{cs2}) are present~\cite{Dima:2020yac}. More precisely, this is because  (\ref{gb2}) and (\ref{cs2})  make a  potential well but not a potential shape with barrier-well-mirror which provides  quasibound states for superradiant instability~\cite{Zouros:1979iw,Arvanitaki:2010sy}.

\section{Discussions }

It is very curious to note that most black holes born from single stars rotate slowly.
We have performed  spontaneous  scalarization  of slowly rotating  black holes
in the  EsGBCS theory. The fairly slow rotating black holes represent the cases with $0<a\le 0.1$~\cite{Fuller:2019sxi}.
 In the slow rotation approximation with $a\ll1$, the GB term is  a larger  term  for Schwarzschild black hole
while the CS term  takes a smaller linear term of $a$-rotation parameter, implying that  its tachyonic instability is determined mainly by the GB term.
The tachyonic instability for slowly rotating  black holes implies  the onset of spontaneous scalarization.

The (2+1)-dimensional  hyperboloidal foliation method is used to show  the tachyonic instability of slowly rotating  black holes when considering a spherically symmetric scalar-mode propagation $u_{00}$. The time  evolution for $\log_{10}|u_{00}(\tau,\rho,\alpha)|$ indicates  stability ($\searrow$), threshold ($\longrightarrow$), and instability ($\nearrow$) with increasing time ($\tau$) shown in Fig. 3.
It is shown  that
slowly rotating black holes are unstable against a spherically symmetric  scalar-mode of $l=m=0$
for  positive coupling $\alpha$ only.  To this end, we have constructed  a threshold  curve   $\alpha=\alpha_{\rm th}(a)$ in Fig. 4  which is the boundary between  stable and unstable black holes
based on the constant scalar modes $[\log_{10}|u_{00}(\tau,\rho,\alpha)|\sim \longrightarrow]$.

 For negative coupling,  there is  no tachyonic instability for scalarization since $\mu^2_{\rm eff}$ has    positive region  outside the horizon.  Additionally,  we could  not find  the $a$-bound for spin-induced  scalarization obtained from the Kerr black hole in the EsGB theory.
However, we propose that the $a$-bound for spin-induced scalarization of Kerr black holes will be shifted from $a\ge0.5$ to a lower bound due to the CS term in the EsGBCS theory with negative coupling~\cite{Zou:2021ybk}.

 \vspace{1cm}

{\bf Acknowledgments}

 This work was supported by the National Research Foundation of Korea (NRF) grant funded by the Korea government (MOE)
 (No. NRF-2017R1A2B4002057).
 
\end{document}